\documentclass[aps,pre,twocolumn,epsfig,superscriptaddress,showpacs,amsmath,amsfonts,amssymb,floatfix]{revtex4}
\usepackage{graphicx}
\usepackage{dcolumn}
\usepackage{bm}

\begin{document}

\title{Generating a Fractal Butterfly Floquet Spectrum in a Class of Driven SU(2) Systems: Eigenstate Statistics}

\author{Jayendra N. Bandyopadhyay}
\affiliation{Department of Physics and Centre for Computational
Science and Engineering, National University of Singapore, Singapore 117542,
Republic of Singapore}
\author{Jiao Wang}
\affiliation{Department of
Physics, and Institute of Theoretical Physics and Astrophysics,
Xiamen University, Xiamen 361005, China}
\author{Jiangbin Gong}
\email[]{phygj@nus.edu.sg}
\affiliation{Department
of Physics and Centre for Computational Science and Engineering,
National University of Singapore, Singapore 117542, Republic of Singapore}
\affiliation{NUS Graduate School for Integrative Sciences and
Engineering, Singapore 117597, Republic of Singapore}
\date{\today}

\begin{abstract}

The Floquet spectra of a class of driven SU(2) systems have been
shown to display butterfly patterns with multifractal
properties. The implication of such critical spectral behavior
for the Floquet eigenstate statistics is studied in this work.
Following the methodologies for understanding the fractal behavior
of energy eigenstates of time-independent systems on the Anderson
transition point, we analyze the distribution profile, the mean
value, and the variance of the logarithm of the inverse
participation ratio of the Floquet eigenstates associated with
multifractal Floquet spectra. The results show that the
Floquet eigenstates also display fractal behavior, but with features
markedly different from those in time-independent
Anderson-transition models.  This motivated us to propose a new type of random unitary
matrix ensemble, called ``power-law random banded unitary
matrix" ensemble, to illuminate the Floquet eigenstate statistics of
critical driven systems.  The results based on the proposed random matrix model are consistent with those obtained from our dynamical examples with or without time-reversal symmetry.
\end{abstract}
\maketitle

\section{\label{sec1} Introduction}

The critical behavior of time-independent systems, especially in
terms of the spectral statistics and the eigenstate statistics,
has attracted great attention. On the spectrum side, Hofstadter's
butterfly spectrum of the Harper model has been a paradigm for
critical spectral statistics, representing a multifractal
spectrum \cite{harper,hofstadter} of a system exactly on the
metal-insulator transition point.
On the eigenstate side, mainly through studies in time-independent
models, such as the power-law random banded matrix (PRBM) model
\cite{PRBM} and the standard Anderson tight-binding model (TBM)
\cite{AndersonModel}, it has been well-established that for a
system on a metal-insulator transition point or the Anderson
transition point \cite{AnderTrans01}, its eigenstates show clear
fractal features. This background of understanding the critical
behavior of time-independent systems motivated our interest in the
critical behavior of periodically driven systems. Below we first
introduce recent related studies of critical Floquet spectra, and
then briefly describe the motivation and the results of this work.

It is well known that the Floquet (quasi-energy) spectrum of a
delta-kicked version of the Harper model also displays Hofstadter's
butterfly patterns \cite{kicked_harper1,kicked_harper2}.
Interestingly, though the kicked Harper model (KHM) can be
classically chaotic, its spectrum, due to its fractal nature, does
not follow the Bohigas-Giannoni-Schmit conjecture \cite{BGS} at all.
This makes the KHM not only a fruitful model for gaining new
insights into the issue of quantum-classical correspondence in
classically chaotic systems, but also an intriguing model to study
critical spectral statistics. Indeed,  for
quite a long time, studies of fractal Floquet spectra were largely
restricted to the KHM and its variants \cite{dana}.  In a
proposal to experimentally realize Hofstadter's butterfly Floquet
spectrum in cold-atom laboratories, Wang and Gong
\cite{JiaoJiangbin1,JMO} recently demonstrated that
Hofstadter's butterfly Floquet spectrum can be synthesized by use of
a double-kicked cold-atom rotor system \cite{DKRM} under a quantum
resonance condition.  Lawton {\it et al.} \cite{JMP} then showed that
the butterfly Floquet spectrum of the cold-atom system studied in Refs.
\cite{JiaoJiangbin1,JMO}
is equivalent to that of the standard KHM if and only if one
system parameter takes irrational values. In addition to motivating
a cold-atom realization of critical Floquet spectra of
periodically driven systems, Refs. \cite{JiaoJiangbin1,JMO} seem to have offered a general strategy for synthesizing critical Floquet spectra in driven systems.

Using an approach extended from Refs. \cite{JiaoJiangbin1,JMO}, recently Wang and Gong \cite{JiaoJiangbin2} showed that the Floquet spectra of a class of driven SU(2) systems also display butterfly
patterns and multifractal properties that are characteristics of highly critical spectra.  This establishes a completely different class of critical driven systems without a connection with the KHM context. Interestingly, the driven SU(2) model in Ref. \cite{JiaoJiangbin2} can be understood as a simple extension of the well-known kicked top model (KTM) \cite{KTM} in the quantum chaos literature. Because the KTM has just been experimentally realized in a cold $^{133}$Cs system \cite{nature}, it can be expected that a critical driven SU(2) system may also be experimentally realized using the collective spin of a $^{133}$Cs atomic ensemble. An alternative experimental realization may be based on a driven two-mode Bose-Einstein condensate
\cite{JiaoJiangbin2,2modeBEC,otherSU2}, which represents a strongly
self-interacting driven system.

Given the above-mentioned class of driven quantum systems with
critical Floquet spectra,  it becomes necessary to study the
behavior of the associated Floquet eigenstates. Theoretically
speaking, because driven SU(2) systems always have a finite number
of Floquet eigenstates, the eigenstate analysis becomes much easier
than in the KHM, with the latter necessarily involving an infinite
number of eigenstates for a fractal Floquet spectrum.  A careful
investigation of the Floquet eigenstates over the entire spectrum
will help to better understand
the critical behavior in time-dependent systems in general.
Experimentally speaking, information about the eigenstate statistics
may be more directly accessible to measurements than a fractal
spectrum.

To analyze the critical behavior of the Floquet eigenstates in driven SU(2) systems, we adopt the same approach as in previous studies of time-independent systems. That is, we shall numerically examine the fluctuations of the eigenstates \cite{QPT}. The eigenstate fluctuations can be
characterized by a set of inverse participation ratios (IPR):
\begin{equation}
P_q^{(\lambda)} = \sum_n |\langle n | \phi_\lambda\rangle|^{2q},
\label{pq}
\end{equation}
where $\lambda$ is the index of the eigenstates, $|\phi_{\lambda}\rangle$ represents one
eigenstate under investigation, and
$\left\{|n\rangle\right\}$ are the basis states. For convenience
we focus on the IPR $P_2$ (i.e., $q=2$). By analogy to critical eigenstate
behavior in time-independent systems,  we expect that $P_2$ scales
anomalously with the Hilbert space dimension $N$ as
\begin{equation}\label{anom_scal}
P_2^{(\lambda)} \sim N^{-D_2^{(\lambda)}},
\end{equation}
where $D_2^{(\lambda)}$ is a fractal dimension of a particular
eigenstate $|\phi_\lambda\rangle$.
But is there also a unique fractal dimension $D_2$ for the average
behavior of all the Floquet eigenstates, for example, via the slope
of the averaged $\ln (P_2)$, denoted $\langle \ln (P_2)\rangle$,
versus $\ln (N)$? To that end, we shall examine if, as the system
gets closer to the thermodynamic limit ($N\rightarrow +\infty$), the
distribution of $\ln(P_2)$ shows signs of a scale-invariant form
\cite{AnderTrans02}. In other words, whether the distribution
function of $\ln (P_2)$, denoted $\Pi [\ln (P_2)]$, only shifts as
$N$ varies.

Certainly, the system under our study has only a finite size $N$.  In
time-independent Anderson-transition studies using the PRBM or the
TBM, it was conjectured that the variance of $\ln (P_2)$, denoted
$\sigma^2(N)$, scales with $N$ as
\begin{equation}
\sigma^2(N) = \sigma^2(\infty) - \frac{A}{N^\gamma},
\label{var_GOE}
\end{equation}
with $\sigma^2(\infty)$, $A$, and $\gamma$ being three adjustable
parameters \cite{cuevas01}. For a $d$-dimensional system on the
Anderson transition point, it was shown that $\gamma$ is related to
$D_2$ by
\begin{equation}
\gamma = \frac{D_2}{2 \beta d},
\label{Dq_GOE}
\end{equation}
where $\beta$ equals $1$ or $2$ depending upon whether or not the
system has time-reversal symmetry \cite{cuevas02}. As one main task
of this work, we shall examine if these results for time-independent
systems still hold for critical Floquet eigenstates.   Furthermore,
we hope to see how the criticality of the eigenstates of unitary
operators differs from the criticality of the eigenstates of
self-adjoint operators.
Results along this direction will also be relevant to recent
investigations on  the ``unitary Anderson model" \cite{unit_Anderson},
the Thue-Morse sequence generating multifractal eigenstates of the
quantum baker's map \cite{arul}, the one-parameter model of quantum
maps showing multifractal eigenstates \cite{Giraud}, as well as
recent experimental and theoretical studies of Anderson transition in
kicked-rotor systems \cite{delande,Jiao}.

We now briefly summarize the main findings of this work. For the
driven SU(2) systems studied here, we consider two different parameter
regimes: in one regime the Floquet spectra display clear butterfly
patterns, and in the other regime, the butterfly patterns of the Floquet spectra have dissolved due to increased strength of the driving fields. For both regimes, we find that $\Pi[\ln (P_2)]$ is not as smooth as observed in the TBM or PRBM, indicating some
non-universal features in dynamical systems. The
$\Pi[\ln (P_2)]$ for cases with dissolved butterfly patterns is however smoother. For either regime, it is found that the ensemble average $\langle\ln (P_2)\rangle$ does
scale linearly with $\ln (N)$, with the slope of the $\langle\ln
(P_2)\rangle$ vs $\ln (N)$ curve clearly defining the fractal dimension
$D_2$ for all the eigenstates.  We also find it possible to fit the variance of $\ln
(P_2)$ by Eq. \eqref{var_GOE}, but with the exponent
$\gamma$ given by
\begin{equation}
\gamma = \frac{D_2}{\beta d} \label{Dq_COE}
\end{equation}
instead (with $d=1$); i.e., a factor of {\it two} is missing from the
denominator as compared with Eq. (4) for time-independent critical
systems. To further understand this difference, we propose a
random matrix model, which we call ``power-law random banded
unitary matrix" (PRBUM) model.
By tuning the parameters of the PRBUM, the $D_2$ value associated with the PRBUM can
be varied.  More interestingly, we observe that the variance $\sigma^2(N)$ of the
PRBUM also follows Eq. (\ref{var_GOE}), with the exponent $\gamma$ again
given by Eq. (\ref{Dq_COE}). This suggests that our findings about
the Floquet eigenstate statistics based on driven SU(2) systems do
reflect some general aspects of critical Floquet eigenstates.

This paper is organized as follows. 
In Sec. \ref{sec3},
we present detailed results of the eigenstates statistics in our driven SU(2) models, with or without time-reversal symmetry.
In Sec. \ref{sec4}, we introduce the PRBUM to represent a class of
critical Floquet operators, discuss the statistics of the
eigenstates of PRBUM, and then compare the associated results with those found in actual dynamical systems.
In Sec. \ref{sec5}, we study the eigenstate statistics of the standard
kicked top model \cite{KTM}, which represents a classically chaotic,
but non-critical, driven system. Section \ref{sec6} concludes this
work.

\section{\label{sec3} Fractal statistics of the Floquet eigenstates in Driven SU(2) Models}

The focus in Ref. \cite{JiaoJiangbin2} is on the fractal spectral statistics. Here, using the same model
we study the statistics of
the Floquet eigenstates.  The first Floquet
operator under study is given by
\begin{equation}
F = \exp\left( i \frac{\eta J_z^2}{2 J} \right) \exp(-i \alpha J_x) \exp\left( - i
\frac{\eta J_z^2}{2 J} \right) \exp(-i \alpha J_x),
\label{Floquet_COE}
\end{equation}
where $J_x, J_y, J_z$ are angular momentum operators satisfying the SU(2) algebra and $J$ is the conserved total angular momentum quantum number that defines a $(2J+1)$-dimensional Hilbert space. Readers can refer to Ref. \cite{JiaoJiangbin2} for detailed descriptions and motivations of this model. This model is also called as a ``double-kicked top model"
(DKTM) in Ref. \cite{JiaoJiangbin2}.

Eigenstates of the $J_z$ operator are denoted as $|m\rangle$, with $J_z|m\rangle = m|m\rangle$. States $\{|m\rangle\}$ will be chosen
as our representation for eigenstate analysis.
To analyze the Floquet eigenstates, it is necessary to express the Floquet operator in symmetric basis
states, a procedure that block-diagonalizes the Floquet matrix.
On the one hand, this will simplify our analysis; on the other hand,
this is necessary for the sake of comparison between an actual
dynamical system and the PRBUM model proposed below.

The DKTM Floquet operator $F$ in Eq. (\ref{Floquet_COE})
has a parity unitary symmetry $R^\dagger F R = F$ where $R=\exp(-i
\pi J_x)$.  This symmetry can be used to block diagonalize the $F$ matrix  into two disconnected sub-matrices associated with either
odd-parity or even-parity subspaces.
Without loss
of generality we only present below results for the $J$-dimensional
odd-parity subspace. Besides the parity symmetry, $F$ also has a
time-reversal anti-unitary symmetry $TFT = F^{\dagger}$, with
\begin{equation}
T = \exp(i \alpha J_x) K,
\end{equation}
where $K$ is the complex conjugation operator. %
To explore the implication of this time-reversal symmetry
for the eigenstate statistics, we shall also consider a variant of $F$, i.e.,
\begin{equation}
F^\prime = \exp\left( i \frac{\eta J_z^2}{2 J} \right) \exp(-i \alpha J_x)
\exp\left( - i \frac{\eta J_z^2}{2 J} \right) \exp(-i \alpha J_y).
\label{Floquet_CUE}
\end{equation}
Evidently, $F'$ differs from $F$ only in the last factor, i.e.,
$\exp(- i \alpha J_x)$ in $F$ is replaced by $\exp(- i \alpha
J_y)$.  Because of this difference, we call $F$ in Eq.
(\ref{Floquet_COE}) the $J_x-J_x$ model and call $F'$ the
$J_x-J_y$ model. It is easy to check that the latter does not have the parity symmetry
or the time-reversal symmetry.  For
the $J_x-J_y$ model, which cannot be reduced to any block-diagonal
form, we examine the eigenstates of the full Floquet matrix.

For both cases we define a dimensionless system parameter,
\begin{equation}
\hbar_\eta \equiv \frac{\eta}{J} = \frac{1}{2}(\sqrt{5}-1) \pi.
\label{parameter1}
\end{equation}
This choice of $\hbar_\eta$ being $\pi$ times
the golden mean is to ensure that the resulting
Floquet eigenstate statistics is indeed representative of
driven systems with fractal Floquet spectra.  As detailed below,
we consider {\it two} different regimes for the product $\alpha J$.
In the first regime defined by $0.95 \leq \alpha J \leq 1.05$, the Floquet spectra show clear butterfly patterns;
in the second regime defined by $9.95 \leq \alpha J \leq 10.05$, the butterfly spectra have dissolved, with fractal dimensions of the spectra increased \cite{JiaoJiangbin2}.

\subsection{\label{sec3subsec1} $J_x-J_x$ model}

This is a time-reversal symmetric system. Because Dyson's circular
ensemble of random unitary matrices \cite{KTM} with time-reversal
symmetry is called ``circular-orthogonal-ensemble" (COE), we
regard the $J_x-J_x$ model as an example of critical COE
statistics.

\subsubsection{$0.95 \leq \alpha J \leq 1.05$}

\begin{figure}[t]
\centering
\includegraphics[width=7.5cm,height=7cm]{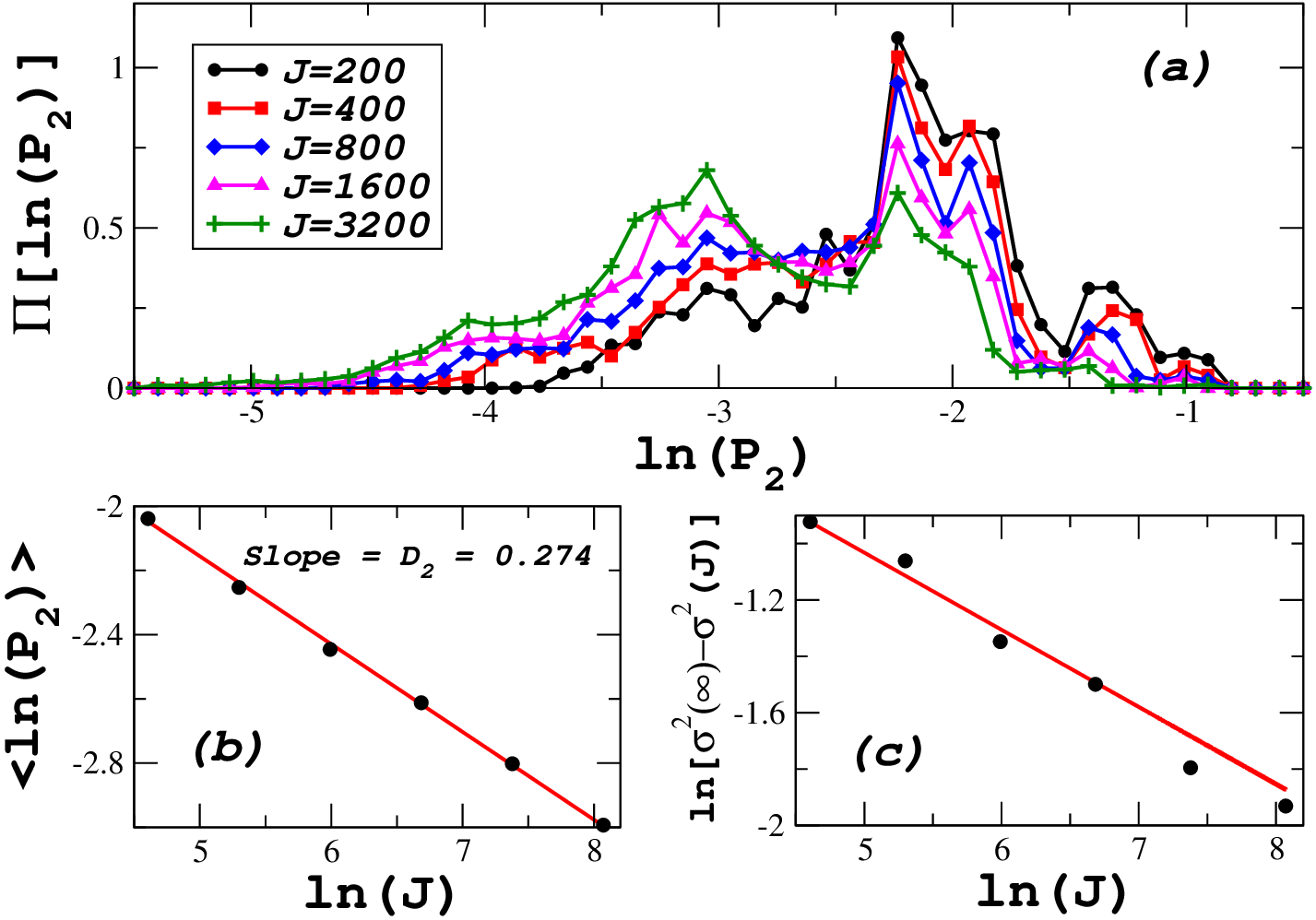}
\caption{(Color online) (a) Distribution of $\ln (P_2)$ for the
$J_x-J_x$ model, with $J = 200, 400, 800, 1600$ and $3200,$
in the representation of odd-parity basis states defined in the text.
The size of the Floquet matrix ensemble is important for numerical
simulation. In order to construct the necessary ensemble, we
consider a range of $\alpha$, i.e., $0.95 \le \alpha J \le 1.05$,
yielding respectively $4000, 2000, 1000, 500$ and $250$
matrices for the different values $J$.  (b) $\langle \ln (P_2) \rangle$, the mean value of $\ln (P_2)$ averaged over all Floquet eigenstates, as a function of $\ln(J)$. The slope of the fitting line gives $D_2 \simeq 0.274$. (c) Logarithm of $[\sigma^2(\infty)-\sigma^2(J)]$ as a function of $\ln(J)$, where $J$ is the dimension of the odd-parity Hilbert subspace.  Filled circles are our numerical
results for the $J_x-J_x$ model, and the solid line is the fitting
of the numerical results using the empirical formula given in Eq.
(\ref{var_GOE}) with  $\sigma^2(\infty) = 0.68$, $A = 1.40$ and
$\gamma = D_2$. The plotted variables here and in all other figures are dimensionless.}
\label{Jx-JxDistAlphaJ1.0}
\end{figure}

Figure \ref{Jx-JxDistAlphaJ1.0}(a) shows the distributions
of the {\it logarithm} of the IPR $P_2$, denoted $\Pi[\ln (P_2)]$,
for different $J$.  It is seen that the distribution function
$\Pi[\ln (P_2)]$ is not as smooth as that observed in early
Anderson-transition studies
\cite{AnderTrans02,cuevas01,cuevas02}. Nevertheless, it is clear
that as $J$ increases, the left tail of $\Pi[\ln (P_2)]$
systematically shifts to the left direction associated
with more negative $\ln (P_2)$. The profile of $\Pi[\ln (P_2)]$,
though somewhat changes as $J$ increases, does maintain its main
features as $J$ increases. Due to these features that are similar
to early findings for the critical eigenstates in time-independent
systems, it can be expected that the average of $\ln (P_2)$ will
show a scaling behavior with $\ln(J)$. As shown in Fig.
\ref{Jx-JxDistAlphaJ1.0}(b), this is indeed the case. Therein, $\langle\ln
(P_2)\rangle$, obtained by averaging $\ln (P_2)$ over all
eigenstates (in the odd-parity subspace), displays an excellent
linear behavior with $\ln (J)$. From the slope of the fitting line in Fig. 1(b), we are able to obtain the fractal dimension $D_2
\simeq 0.274$.

The distribution profile $\Pi[\ln (P_2)]$ in Fig. 1(a) is seen to display rich features, with significant fluctuations and multiple notable peaks.
Qualitatively, this reflects that our system is an actual dynamical
system and hence the underlying rich dynamics will manifest itself
through some non-universal statistical features. Related to this observation
we also note that
in our calculations, all the Floquet eigenstates are
treated equally and all of them are used for averaging. This is in
contrast to the common procedure in analyzing time-independent
critical systems, where only those energy eigenstates in a certain
small energy window around zero eigenvalue are included to examine
the distribution of $\ln (P_2)$
\cite{AnderTrans02,cuevas01,cuevas02}.  The justification for
including all Floquet states in our analysis is as follows:  the
quasi-energy spectra lie on a unit circle and hence all states
with different eigenphases on the unit circle should be treated on
equal footing.  To double check this understanding, we have also taken
windows of different widths centered around {\it zero} value of
the eigenphase and then calculate the distribution of $\ln
(P_2)$. No improvement in the smoothness of $\Pi(\ln P_2)$ is
found.  Rather, we obtained similar distribution of $\ln (P_2)$
with clear fluctuations.  It is also tempting to connect the non-universal features of
$\Pi(\ln P_2)$ with the phase space structures of the underlying classical limit.
However, such a perspective, which calls for a good understanding of quantum-classical correspondence in critical systems, is unlikely to succeed
because the classical limit of our dynamical model is completely chaotic \cite{JiaoJiangbin2}.

In Fig. \ref{Jx-JxDistAlphaJ1.0}(c), we plot $\ln[\sigma^2(\infty)-\sigma^2(J)]$ vs $\ln(J)$ (filled circles), where $\sigma^2(J)$ is the variance of
$\ln(P_2)$ and $\sigma^2(\infty)$ is a fitting parameter, whose value is found by fitting our data points
with the empirical formula given in Eq. (\ref{var_GOE}). As seen
in Fig.~\ref{Jx-JxDistAlphaJ1.0}(c), the fitting is reasonably good,
yielding that $[\sigma^2(\infty)-\sigma^2(J)]$ scales as
$J^{-\gamma}$, with $\gamma=D_2$ [$D_2$ is already determined by the fitting in Fig.
\ref{Jx-JxDistAlphaJ1.0}(b)], $\sigma^2(\infty) \simeq 0.68$,
and $A \simeq 1.40$.  Despite obvious fluctuations around the fitting curve,
the result in Fig. \ref{Jx-JxDistAlphaJ1.0}(c) suggests that the tool borrowed from traditional Anderson-transition studies for time-independent systems can be still useful here.
Furthermore (probably more interestingly), the fitting in Fig. \ref{Jx-JxDistAlphaJ1.0}(c) also unexpectedly reveals a big difference from
what can be expected from Eq. (\ref{Dq_GOE}) with $d=1$ and
$\beta=1$: here $\gamma=D_2$ instead of $D_2/2$.  Therefore, an intriguing difference between time-independent critical systems and periodically driven critical systems is observed here.

\subsubsection{$9.95 \leq \alpha J \leq 10.05$}

\begin{figure}[t]
\centering
\includegraphics[width=7.5cm,height=7cm]{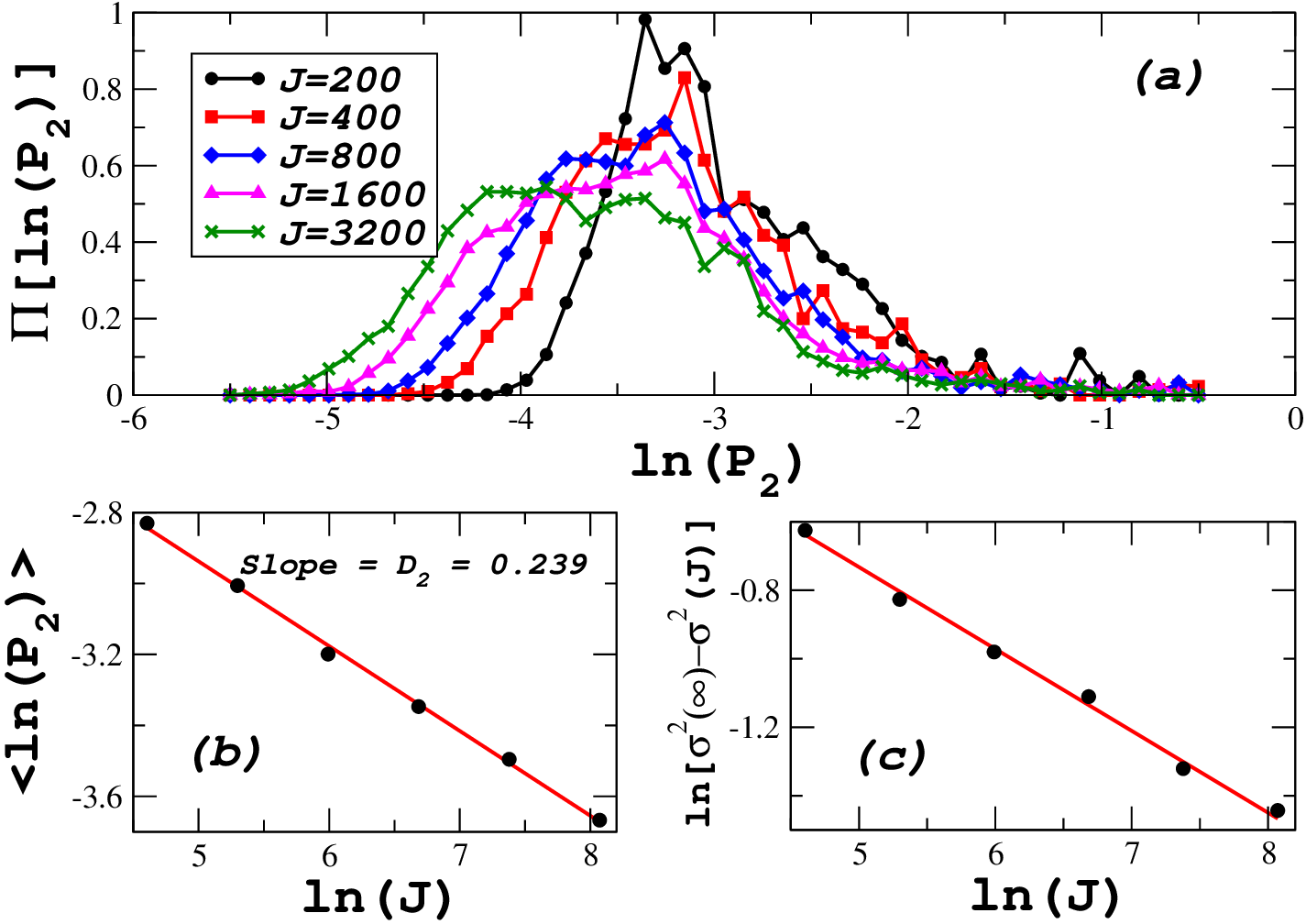}
\caption{(Color online) (a) Distribution of $\ln(P_2)$ for the $J_x-J_x$ model
with $9.95 \leq \alpha J \leq 10.05$ is presented for the odd-parity subspace.
Other parameters are the same as in Fig. \ref{Jx-JxDistAlphaJ1.0}. (b) $\langle
\ln (P_2) \rangle$ is plotted as a function of $\ln(J)$. The slope of the
fitting line gives $D_2 \simeq 0.239$. (c) Logarithm of $[\sigma^2(\infty)-\sigma^2(J)]$ as a function
of $\ln(J)$, where $J$ is the dimension of the odd-parity Hilbert subspace.
Filled circles are our numerical results for the $J_x-J_x$ model, and the solid line
is the fitting
of the numerical results using the empirical formula given in Eq.
(\ref{var_GOE}), with  $\sigma^2(\infty) \simeq 0.77$, $A = 1.59$ and
$\gamma = D_2$.}
\label{Jx-JxDistAlphaJ10.0}
\end{figure}

As mentioned above, for this parameter regime the butterfly patterns in the Floquet spectra have dissolved almost completely.  We present the associated eigenstate statistics in Fig. \ref{Jx-JxDistAlphaJ10.0}.
In Fig. \ref{Jx-JxDistAlphaJ10.0}(a), we show the distribution profile of $\ln (P_2)$ for different $J$. In contrast to the previous case shown in Fig. 1(a), $\Pi[\ln(P_2)]$ is now much smoother (essentially only one peak is left). From the same panel, we also see a systematic left-shift of the distribution function as $J$ increases. This systematic left-shift leads to an evident linear behavior of the average value of $\ln (P_2)$ as a function of $\ln (J)$, as shown in  Fig. \ref{Jx-JxDistAlphaJ10.0}(b).  The slope of the fitting line in Fig. \ref{Jx-JxDistAlphaJ10.0}(b) gives the fractal dimension $D_2 \simeq 0.256$.  Comparing this result with that in Fig. 1(b), one sees that though the fractal dimension of the Floquet spectra increases due to increasing $\alpha J$ \cite{JiaoJiangbin2}, the fractal dimension of the associated eigenstates may decrease.

In Fig.~\ref{Jx-JxDistAlphaJ10.0}(c), we examine the variance of $\ln (P_2)$ as a function of $\ln(J)$ (again, for the odd-parity subspace).  Same as in Fig. 1(c), we fit our results with the empirical formula given in Eq. (\ref{var_GOE}). The fitting in Fig. \ref{Jx-JxDistAlphaJ10.0}(c) is better than that in Fig. 1(c), consistent with the fact that the distribution of $\ln (P_2)$ is quite smooth here.  The fitting in Fig. \ref{Jx-JxDistAlphaJ10.0}(c) gives
$\sigma^2(\infty) \simeq 0.77, A \simeq 1.59$, and $\gamma = D_2$, where the value of $D_2$ is found in Fig. \ref{Jx-JxDistAlphaJ10.0}(b).  The finding that $\gamma$ is not equal to $D_2/2$ but $D_2$ again
strengthens our early observation from Fig. 1.

\subsection{\label{sec3subsec2} $J_x-J_y$ model}

\begin{figure}[b]
\centering
\includegraphics[width=8cm,height=7.5cm]{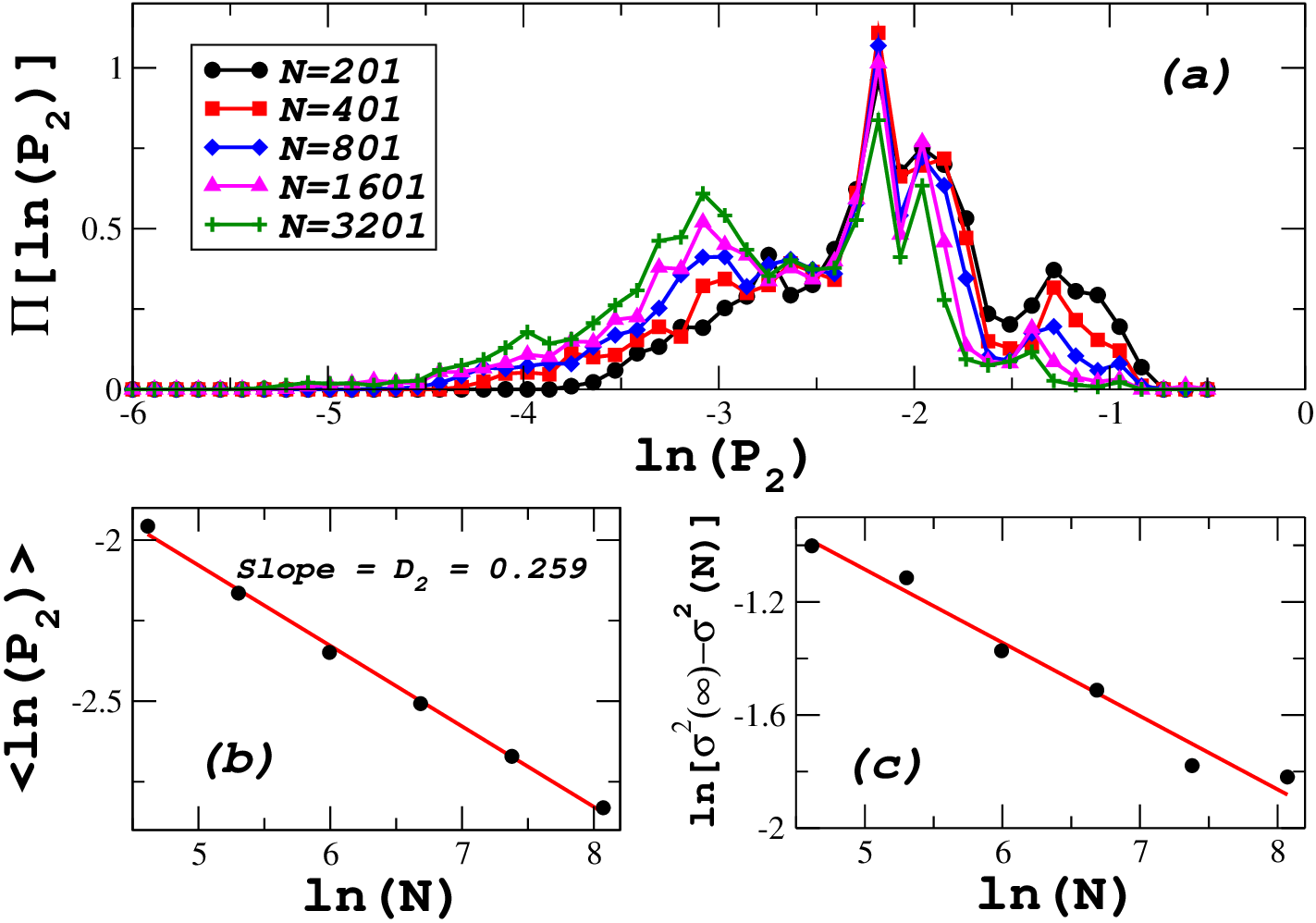}
\caption{(Color online) (a) Distribution of $\ln (P_2)$ for the
$J_x-J_y$ model, with $J = 100 (4000), 200 (2000), 400 (1000), 800
(500),$ and $1600 (250)$. The numbers in the brackets are the size
of the Floquet matrix ensemble. The dimension of the Hilbert
space is given by $N=2J+1$. (b) $\langle \ln (P_2) \rangle$, the
ensemble mean value of $\ln (P_2)$ averaged over all Floquet
eigenstates, as a function of $\ln (N)$.  The slope of the curve
of $\langle \ln (P_2) \rangle$ vs $\ln (N)$ gives $D_2 \simeq
0.259$. (c) Logarithm of $[\sigma^2(\infty)-\sigma^2(N)]$ vs $\ln(N)$.
Filled circles are
numerical results for the $J_x-J_y$ model, and the solid line is
the fitting of the numerical results using the empirical formula
given in Eq. (\ref{var_GOE}), with $\sigma^2(\infty) = 0.92$, $A =
1.04$, and $\gamma = D_2/2$.}
\label{Jx-JyDistAlphaJ1.0}
\end{figure}

To verify if our findings above are general, we now turn
to the $J_x-J_y$ model [Eq. (\ref{Floquet_CUE})].  Due to the lack of time-reversal symmetry
here, this case can be regarded as an example of critical
``circular-unitary-ensemble" (CUE) statistics.  All the eigenstates of the Floquet operator $F'$ will be
considered.

\subsubsection{$0.95 \leq \alpha J \leq 1.05$}

For this regime where the butterfly patterns of the Floquet spectra can be clearly seen, Fig. \ref{Jx-JyDistAlphaJ1.0}(a) displays the distribution of $\ln (P_2)$ for different Hilbert space dimension $N=2J+1$.  Analogous to the previous case with time-reversal symmetry,
$\Pi[\ln (P_2)]$ displays interesting fluctuations. As $N$
increases, $\Pi[\ln (P_2)]$ undergoes changes in its profile, shifts
its left tail, but also maintains many features. In Fig.
\ref{Jx-JyDistAlphaJ1.0}(b) we obtain again a nice linear scaling behavior of $\langle \ln (P_2) \rangle$ with $\ln (N)$.  From the slope of the linear scaling, we obtain the fractal dimension $D_2 \simeq 0.259$.  This $D_2$ value is different from that for the $J_x-J_x$
model with the same values of $\alpha J$(Note that the spectral statistics for the $J_x-J_y$ model also differs from that for the $J_x-J_x$ model \cite{JiaoJiangbin2}).

Same as in Fig. \ref{Jx-JxDistAlphaJ1.0}(c), in Fig.
\ref{Jx-JyDistAlphaJ1.0}(c) we study the variance of $\ln (P_2)$ [now denoted $\sigma^2(N)$] as a
function of $\ln(N)$, using the fitting formula
given in Eq. (\ref{var_GOE}). The fitting, though with clear fluctuations, yields that $[\sigma^2(\infty)-\sigma^2(N)]$ scales as
$N^{-\gamma}$, with $\sigma^2 (\infty) \simeq 0.92$, $A \simeq
1.04$, and $\gamma=D_2/2$ [$D_2$ value obtained from Fig.
\ref{Jx-JyDistAlphaJ1.0}(b)].  Remarkably, though Eq. (4) with $d=1$ and
$\beta=2$ (because of the lack of time-reversal symmetry) predicts
$\gamma=D_2/4$, here we have $\gamma=D_2/2$ instead.  The
important common feature shared by the $J_x-J_y$ model and the
$J_x-J_x$ model is thus the missing of a factor of 2 in the
numerically obtained $\gamma$ value as compared with the empirical
formula for time-independent critical systems. This interesting finding also supports the use of Eq. (\ref{var_GOE})
as a tool for understanding Floquet eigenstate statistics.
Our numerical observations here will be further strengthened by
a random matrix model.

\subsubsection{$9.95 \leq \alpha J \leq 10.05$}

\begin{figure}[b]
\centering
\includegraphics[width=7.5cm,height=7cm]{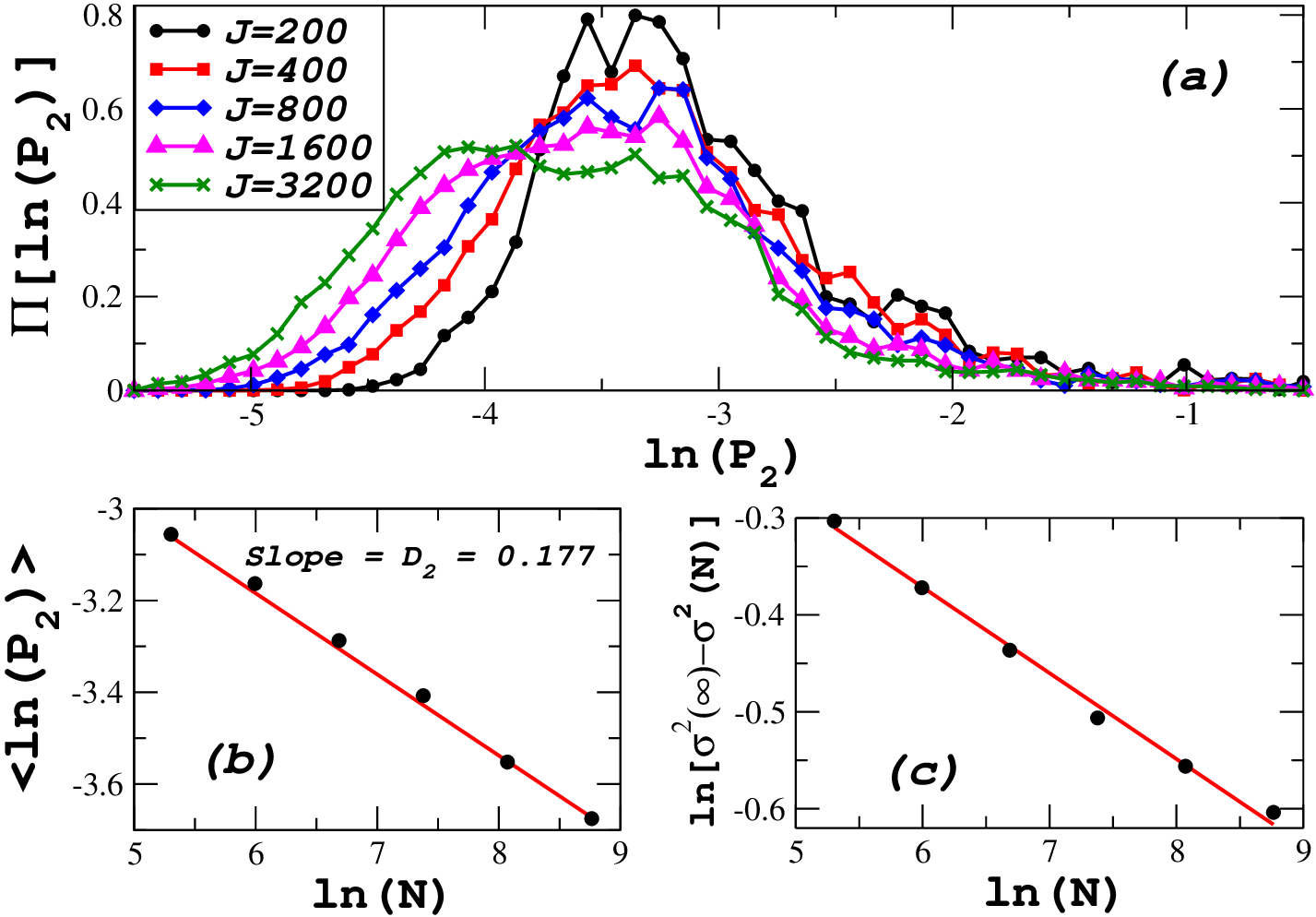}
\caption{(Color online) (a) Distribution of $\ln(P_2)$ for the $J_x-J_y$ model
with $9.95 \leq \alpha J \leq 10.05$ is presented for different Hilbert space
dimension $N=2J+1$. Other parameters are the same as in Fig. \ref{Jx-JyDistAlphaJ1.0}.
(b) $\langle \ln (P_2) \rangle$ is plotted as a function of $\ln(N)$. The slope
of this linear curve gives $D_2 \simeq 0.177$. (c) Logarithm of $[\sigma^2(\infty)-\sigma^2(N)]$ vs $\ln(N)$.
Filled circles are our numerical results for the $J_x-J_y$
model, and the solid line is the fitting of the numerical results using the
empirical formula given in Eq. (\ref{var_GOE}), with  $\sigma^2(\infty) \simeq
1.11$, $A = 1.17$ and $\gamma = D_2/2$.}
\label{Jx-JyDistAlphaJ10.0}
\end{figure}

Just like the $J_x-J_x$ model, in this regime the butterfly patterns of the Floquet spectra have dissolved.
The statistical properties of
the Floquet eigenstates are shown in Fig.~\ref{Jx-JyDistAlphaJ10.0}. In Fig.~\ref{Jx-JyDistAlphaJ10.0}(a), the distributions of $\ln (P_2)$ is seen to be much smoother than those seen in Fig. 3(a).  This is somewhat expected from our early findings in the $J_x-J_x$ model.
Figure \ref{Jx-JyDistAlphaJ10.0}(b) shows a linear scaling of $\langle \ln (P_2)\rangle$ vs $\ln(N)$, with its slope
giving $D_2 \simeq 0.177$. In Fig. \ref{Jx-JyDistAlphaJ10.0}(c), we study the variance of $\ln (P_2)$
as a function of $\ln(N)$, as compared with the empirical formula given in Eq. (\ref{var_GOE}):
the fitting with the empirical formula is excellent, yielding
$\sigma^2(\infty) \simeq 1.11, A \simeq 1.17$, and $\gamma = D_2/2$, where
the value of $D_2$ is determined in Fig. \ref{Jx-JyDistAlphaJ10.0}(b).  Once again, here we find
$\gamma = D_2/2$ instead of $\gamma = D_2/4$ [as suggested by Eq. (4) with $\beta=2$].

\section{\label{sec4} Eigenstate Statistics of PRBUM}

In studies of time-independent critical systems, the PRBM model at
criticality \cite{PRBM} has proved to be fruitful. The PRBM is an
ensemble of random Hermitian matrices whose matrix elements
$\{H_{ij}\}$ are independently distributed Gaussian random numbers
with mean $\langle H_{ij} \rangle = 0$ and the variance satisfying
\begin{equation}
\sigma^2 (H_{ij}) = \left[ 1 + \left(\frac{|i-j|}{b}\right)^{2 g}
\right]^{-1}.
\label{VarH_ij}
\end{equation}
The case $g = 1$ represents the critical point and $0 < b <
\infty$ is a parameter characterizing the ensemble. A
straightforward interpretation of this model is that it describes
a one-dimensional sample with random long-range hopping, with the
hopping amplitude decaying as $|i-j|^{-1}$.  Motivated by our results
above for critical Floquet states, we aim to propose a class of
random unitary matrices, whose Floquet eigenstate statistics can
show some general aspects of critical statistics and can be used to
shed some light on actual dynamical systems.  Our natural starting point for
generating such random unitary matrices are the Hermitian PRBM.

\subsection{\label{sec4subsec1} Algorithm}

\begin{figure}[t]
\centering
\includegraphics[width=7.5cm,height=6.3cm]{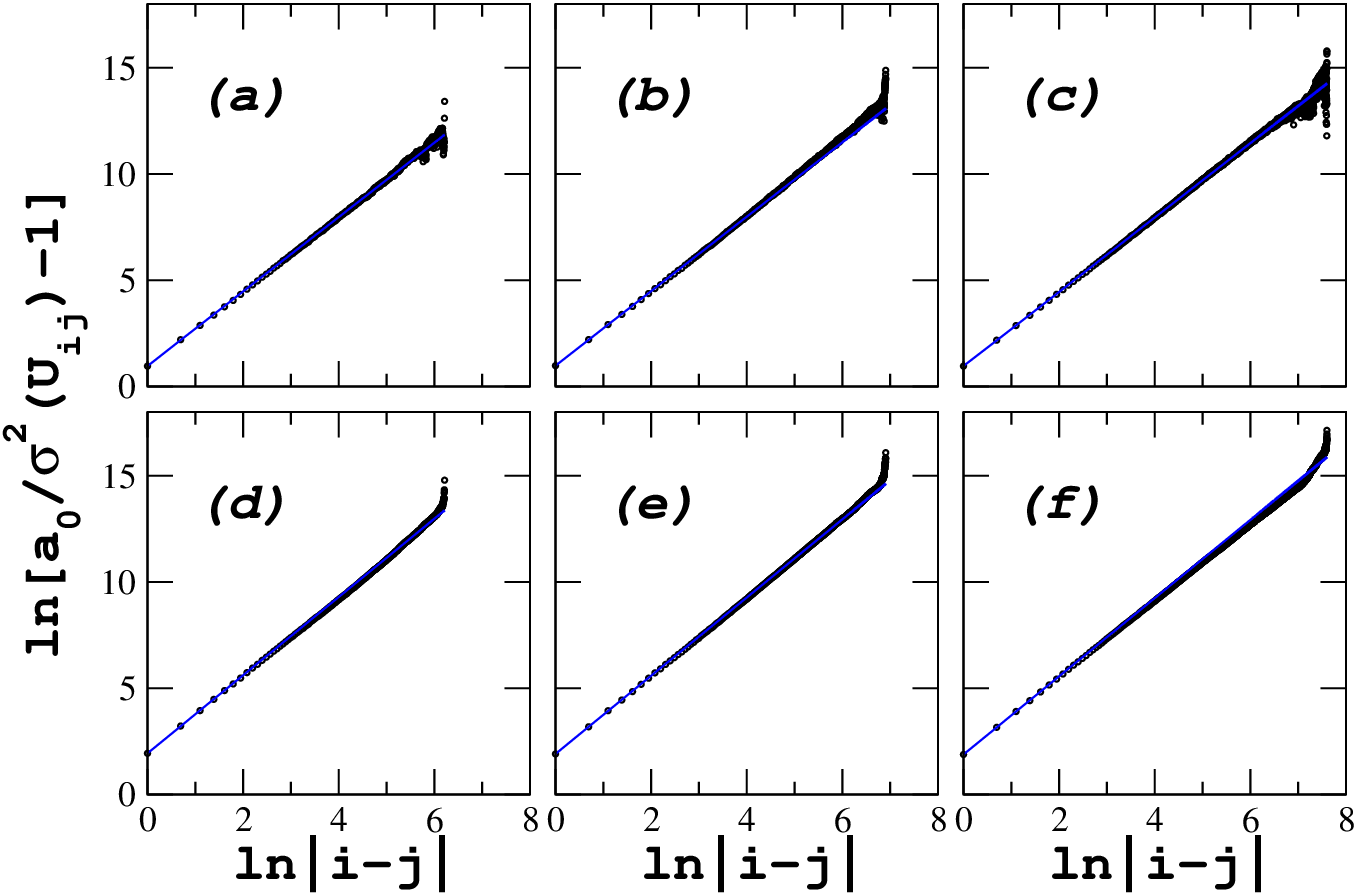}
\caption{(Color online) The variance of the matrix elements of the random unitary matrices generated by
Mazzadri's algorithm with PRBM as the input. To demonstrate the power law scaling,
the dependence of $\ln(a_0/\sigma^{2}(U_{ij})-1)$ on $\ln|i-j|$ is plotted, where $U_{ij}$ represents
a matrix element at the $i$th row and $j$th column.
Here
the PRBM ensemble as the input is set at the critical point
$g = 1.0$ with the parameter $b = 0.1$. Panels (a)-(c) are
for PRBUM-COE, with the dimension $N = 500, 1000,$
and $2000$, respectively.  The fitting function of Eq. (\ref{var_unitary}) (solid lines) gives $a_0 = 0.165 \pm 0.004$,
$b_0 = 0.575 \pm 0.003$, and $g_0 = 0.875 \pm 0.002$.
Panels (d)-(f) are for PRBUM-CUE, with the dimension $N = 500,
1000,$ and $2000$, respectively. The fitting function of Eq. (\ref{var_unitary}) (solid lines)
yields $a_0 = 0.277 \pm 0.005$, $b_0 = 0.355 \pm 0.004$, and
$g_0 = 0.921 \pm 0.001$.}
\label{var_prbum}
\end{figure}

To generate a random unitary matrix from a Hermitian matrix in the PRBM
ensemble, we employ the algorithm by Mezzadri, whose original motivation is to
generate CUE random matrices \cite{Mezzadri} from general Gaussian
random matrices.  For the sake of completeness, we have presented a
description of Mezzadri's algorithm in Appendix \ref{App1}. For our purpose, that is, to
generate a critical random unitary matrix, we first set the starting point
of Mezzadri's algorithm as a PRBM ensemble at the critical point ($g = 1.0$).
We then generate an ensemble of random unitary
matrices (denoted $U$) of the CUE class.  Significantly, because of the use of PRBM as the input for
Mezzadri's algorithm, we find that the variance of the matrix elements $\{U_{ij}\}$ thus
obtained also satisfies a power-law, i.e.,
\begin{equation}
\sigma^2(U_{ij}) = a_0 \left[ 1 + \left(\frac{|i-j|}{b_0}\right)^{2 g_0}
\right]^{-1}.
\label{var_unitary}
\end{equation}
Here the parameter $a_0$ is a common prefactor of the matrix
elements, which can be determined by the unitary condition. The
parameters $g_0$ and $b_0$ in Eq. (\ref{var_unitary}) depend on
the parameters $g$ and $b$ of the PRBM used. As three computational examples,
panels (d)-(f) of Fig.
\ref{var_prbum} present the dependence of $\ln[a_0/\sigma^{2}(U_{ij})-1]$ upon $\ln|i-j|$,
for three ensembles of random unitary matrices we generated, with sizes $N=500, 1000$, and $2000$. If the scaling of $\sigma^{2}(U_{ij})$ with $|i-j|$ is indeed a power law as described by Eq. (\ref{var_unitary}), then one should see a linear dependence of $\ln[a_0/\sigma^{2}(U_{ij})-1]$ in $\ln|i-j|$.
This is indeed the case in Figs.~\ref{var_prbum}(d)-(f).  Note that the deviations in Figs.~\ref{var_prbum}(d)-(f) from the fitting straight lines at very large values of $\ln|i-j|$ are due to two trivial reasons. First, for very large $|i-j|$, the value of $\sigma^{2}(U_{ij})$ is vanishingly small and hence $\ln[1/\sigma^{2}(U_{ij})-1]$ becomes extremely large, thus yielding large fluctuations. Second and more importantly, for a fixed matrix size, if $|i-j|$ is very large, then the available number of matrix elements become insufficient for good statistics. Indeed, as the matrix size increases from $N=500$ to $N=2000$, it is seen from Figs.
\ref{var_prbum}(d)-(f) that the validity window of the linear fitting gradually extends to larger values of $\ln|i-j|$.

The random unitary matrices generated in the above manner, with their  matrix elements satisfying the power-law scaling of Eq. (\ref{var_unitary}), are defined as ``power-law random banded unitary matrix" of the CUE type (PRBUM-CUE). As detailed in Appendix A, one can then generate PRBUM of the COE type (PRBUM-COE) via $V = U U^T$. As shown in panels (a)-(c) of Fig. \ref{var_prbum}, the variance of the matrix elements of PRBUM-COE  also obeys Eq. (\ref{var_unitary}), with different values of $g_0$ and $b_0$.

To check whether
the PRBUM-COE and PRBUM-CUE ensembles show critical statistics,
we analyzed their eigenstates, especially in terms of the
distribution and the scaling of $\ln (P_2)$.  It is found that as we
tune the parameter $b$ of the PRBM used in the algorithm, the
resulting fractal dimensions $D_2$ can be also tuned continuously. For example, the $D_2$ value of PRBUM
can be made close to that of
our driven SU(2) models. In particular, at $b = 0.1$, we obtain
$g_0 \simeq 0.92$ for PRBUM-COE  and $g_0 \simeq 0.88$ for
PRBUM-CUE, yielding $D_2
\simeq 0.279$ and  $D_2 \simeq
0.251$, respectively. These two $D_2$ values are quite close to the $D_2$ values of the $J_x-J_x$
and $J_x-J_y$ models found in Fig. 1 and Fig. 3.  Below we describe these findings in
detail.

\subsection{\label{sec4subsec2} PRBUM-COE}

\begin{figure}[t]
\centering
\includegraphics[width=7.5cm,height=7cm]{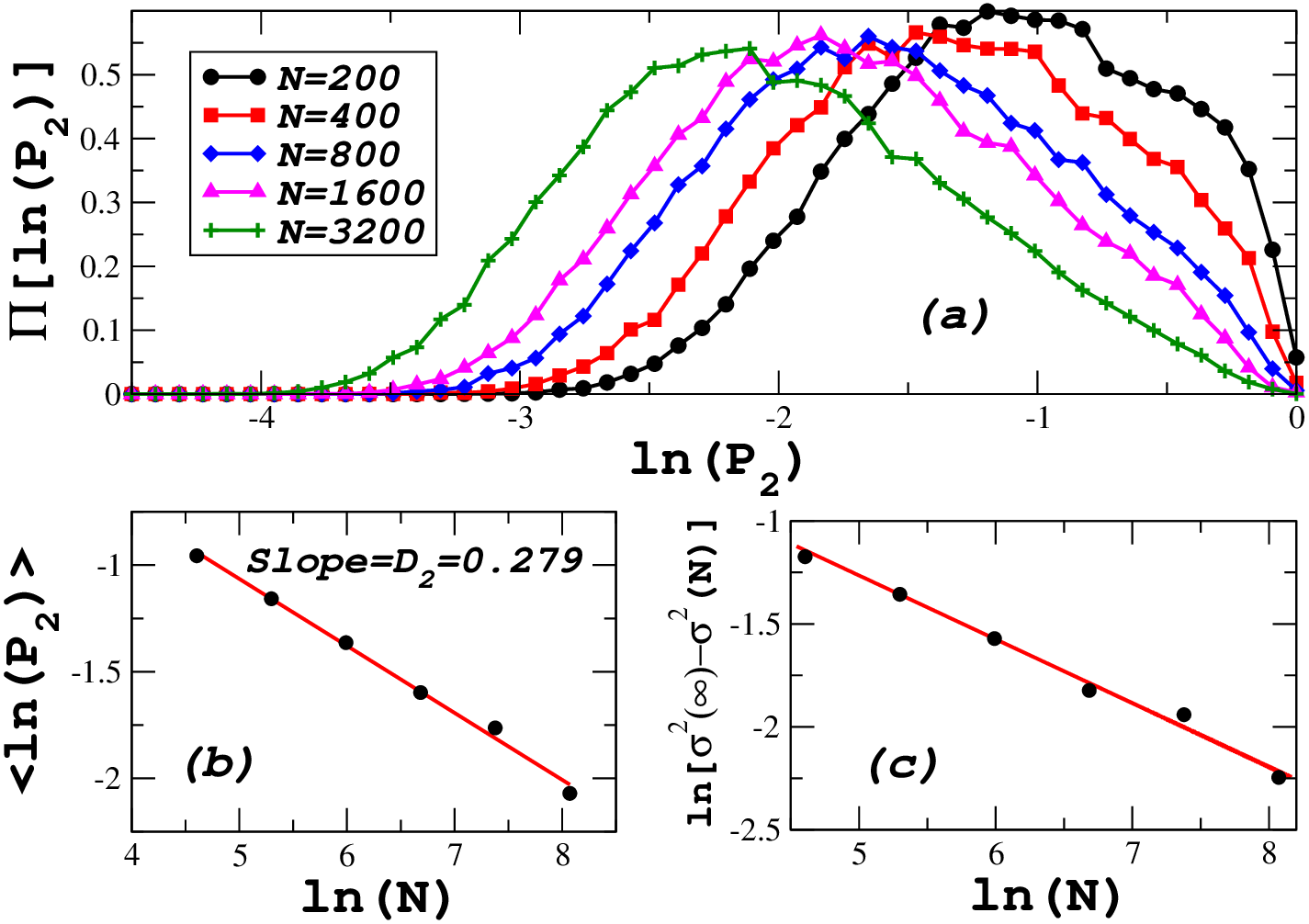}
\caption{(Color online) (a) Distribution of $\ln (P_2)$ obtained for
PRBUM-COE, with the matrix dimension $N = 200 (4000), 400 (2000),
800 (1000), 1600 (500)$ and $3200 (250)$.  The numbers in the
brackets give the size of the ensemble. (b) Same as in Fig.
\ref{Jx-JxDistAlphaJ1.0}(b) and Fig. 2(b), yielding $D_2 \simeq 0.279$. (c) Same as in
Fig. \ref{Jx-JxDistAlphaJ1.0}(c) and Fig. 2(c), but with $J$ replaced by $N$. The fitting
curve gives $\sigma^2(\infty) \simeq 0.60$, $A \simeq 1.33$, and
$\gamma=D_2$.}
\label{COE_Dist}
\end{figure}

This random unitary matrix ensemble is intended to model a critical
Floquet operator with time-reversal symmetry. The results for PRBUM-COE generated from PRBM with $b=0.1$ are shown in
Fig. \ref{COE_Dist}.  In Fig.
\ref{COE_Dist}(a), we show the distributions of $\ln (P_2)$ for
different values of the matrix dimension $N$ (which is the
counterpart of $J$ in the $J_x-J_x$ model), with all the eigenstates of the PRBUM-COE ensemble  considered.  In contrast to the
$J_x-J_x$ dynamical model with a small $\alpha J$ [see Fig. 1(a)], $\Pi[\ln(P_2)]$ here displays very smooth
behavior.  Figure
\ref{COE_Dist}(b) depicts a nice linear relation between $\langle
\ln P_2 \rangle$ and $\ln (N)$. The
slope of the straight line in Fig. \ref{COE_Dist}(b) gives the fractal
dimension $D_2 \simeq 0.279$, a value close to that in the $J_x-J_x$ model with $0.95\leq \alpha J\leq 1.05$.
As in Fig.~\ref{Jx-JxDistAlphaJ1.0}(c), Fig.
\ref{COE_Dist}(c) shows the fitting of the variance of $\ln (P_2)$
with $N$, using Eq. (\ref{var_GOE}). Interestingly, the values of the
fitting parameters are found to be $\sigma^2 (\infty) \simeq 0.60$,
$A=1.33$, both are similar to those determined in Fig. 1(c). More interestingly, this fitting shows that
$[\sigma^2(\infty)-\sigma^2(N)]$ scales as $N^{-\gamma}$, with
$\gamma=D_2$. This supports our finding in Fig.
\ref{Jx-JxDistAlphaJ1.0}(c) and Fig. 2(c).
We have also studied other cases of PRBUM-COE using other PRBM as the input of Mezzadri's algorithm. For example, we find that
if the parameter $b$ is set at $\sim 0.08$, then the $D_2$ of the PRBUM-COE ensemble is around 0.24, which is close to the
$D_2$ value previously found in the $J_x-J_x$ model with $9.95\leq \alpha J\leq 10.05$.
These results clearly support our use of PRBUM-COE to
illuminate the critical eigenstate statistics in the $J_x-J_x$ model.

\subsection{\label{sec4subsec3} PRBUM-CUE}

\begin{figure}[t]
\centering
\includegraphics[width=7.5cm,height=7cm]{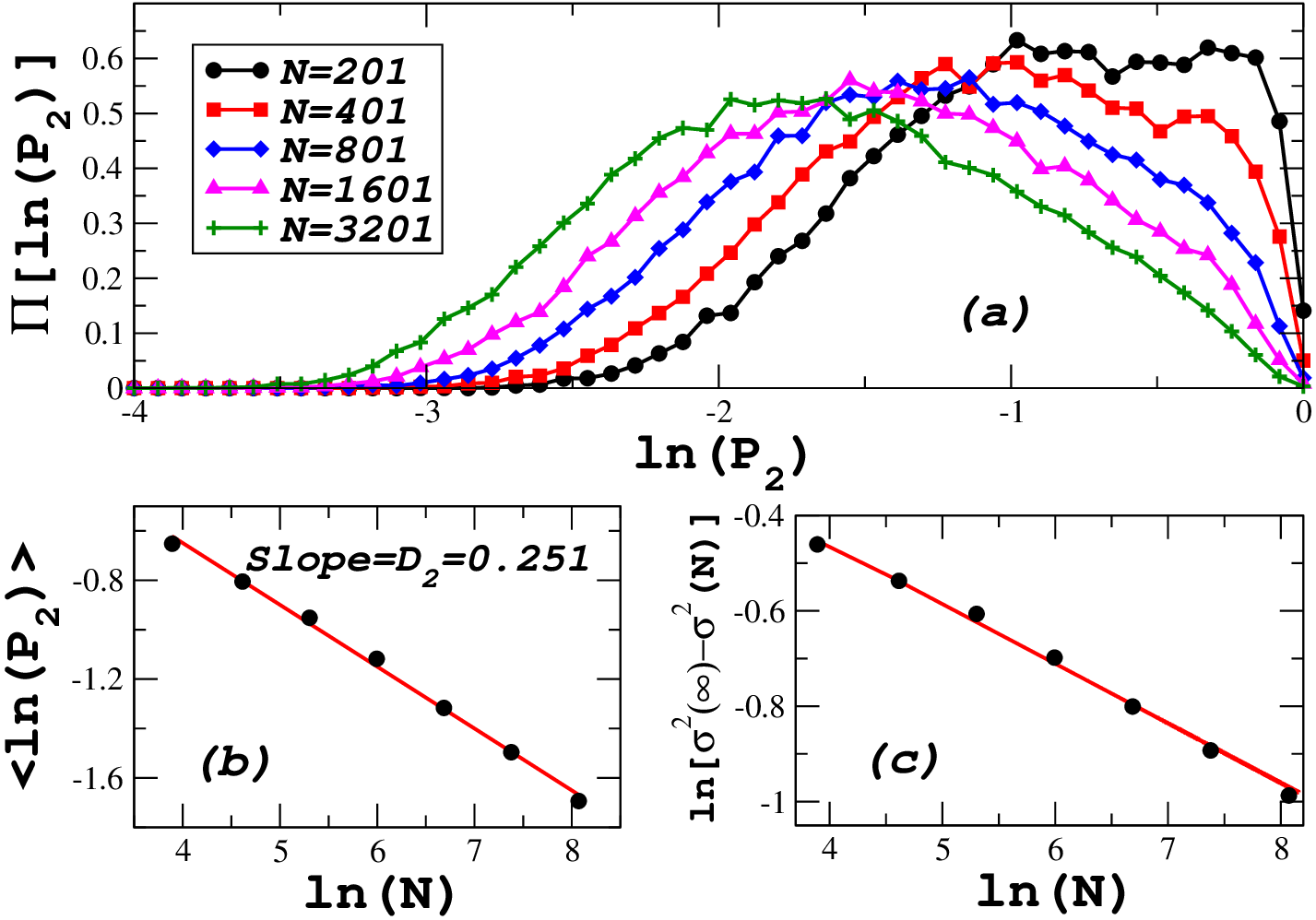}
\caption{(Color online) (a) Distribution of $\ln (P_2)$ obtained for
PRBUM-CUE, with the matrix dimension $N = 201 (4000), 401 (2000),
801 (1000), 1601 (500)$ and $3201 (250)$. The numbers in the
brackets give the size of the ensemble. (b) Same as in Fig.
\ref{Jx-JyDistAlphaJ1.0}(b) and Fig. 4(b), yielding  $D_2 \simeq 0.251$. (c) Same as in
Fig. \ref{Jx-JyDistAlphaJ1.0}(c) and Fig. 4(c), the fitting gives $\sigma^2(\infty) \simeq
0.85$, $A \simeq 1.05$, and $\gamma=D_2/2$.}
\label{CUE_Dist}
\end{figure}

This ensemble aims to model a critical Floquet operator without
time-reversal symmetry.   All eigenstates of
an ensemble of PRBUM-CUE matrices are used for our statistical
analysis. For $b=0.1$, Fig. \ref{CUE_Dist}(a) displays $\Pi[\ln (P_2)]$ versus
$\ln (P_2)$, showing again a smooth dependence. Figure
\ref{CUE_Dist}(b) shows the corresponding $\langle \ln (P_2) \rangle$ versus $\ln
(N)$, which yields the fractal dimension $D_2 \simeq 0.251$. In
Fig. \ref{CUE_Dist}(c), we fit the dependence of $\ln[\sigma^2(\infty)-\sigma^2(N)]$ in $\ln(N)$, yielding
$[\sigma^2(\infty)-\sigma^2(N)] \sim N^{-\gamma}$, with
$\gamma=D_2/2$ [instead of $D_2/4$ predicted by Eq.
(4)]. This also confirms our early observations in the
$J_x-J_y$ model. The values of the fitting parameters are found to
be $\sigma^2(\infty) \simeq 0.85$ and $A \simeq 1.05$, which are close
to what we found in Fig. \ref{Jx-JyDistAlphaJ1.0}(c).  We have also checked that if we perform analogous calculations for $b\sim 0.07$,
then the $D_2$ value for the PRBUM-CUE ensemble will be close to that found in Fig. 4(b).
Given these results, we are led to the
conclusion that PRBUM as proposed above do share some general aspects with periodically driven systems having critical
eigenstate statistics.

\section{\label{sec5}  Floquet eigenstate statistics of the kicked top model}

\begin{figure}[b]
\centering
\includegraphics[width=8cm,height=7cm]{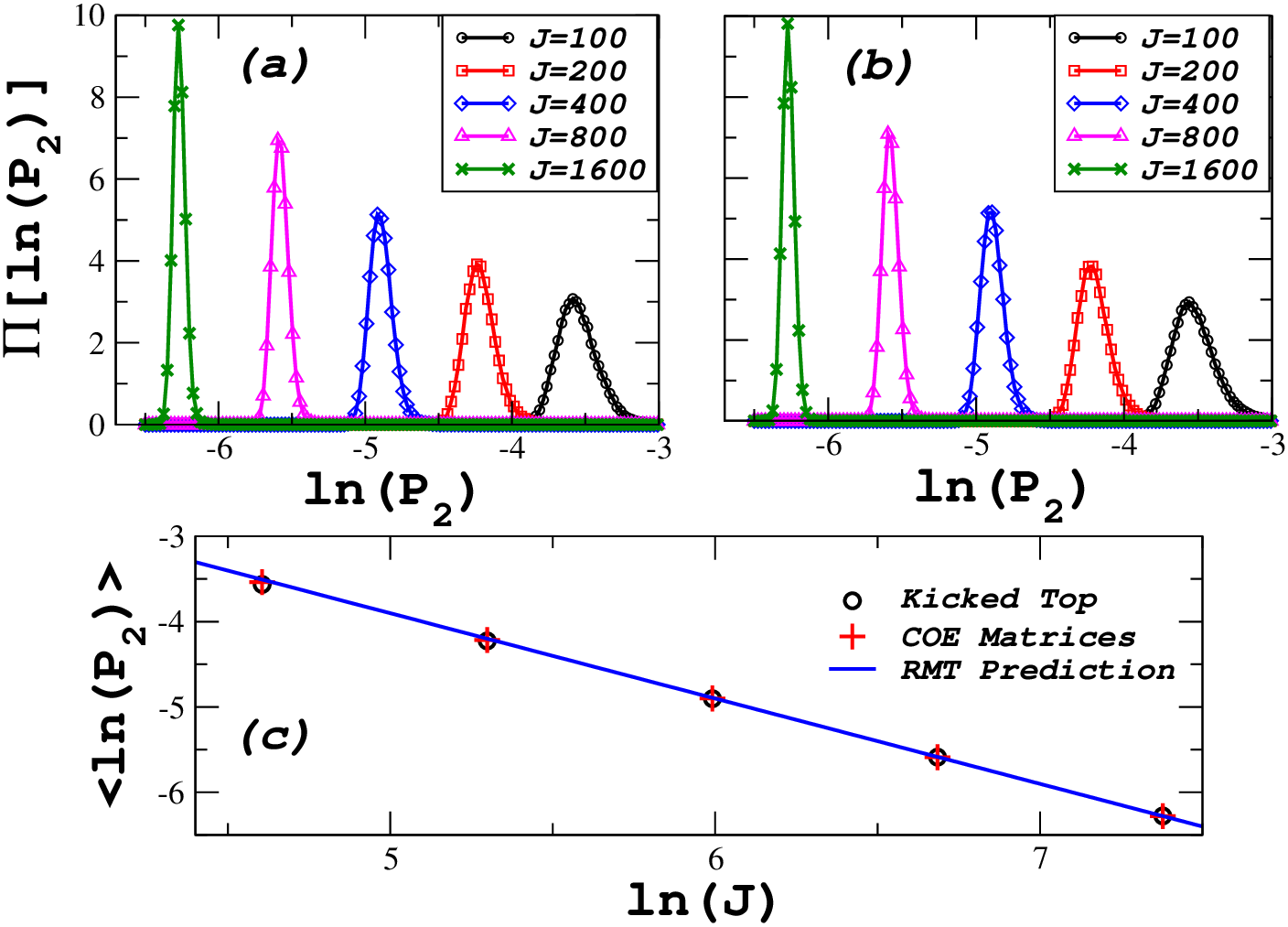}
\caption{(Color online)(a) Distributions of $\ln (P_2)$ for the
standard classically chaotic kicked top model, for $J = 100 (4000),
200 (2000), 400 (1000), 800 (500)$ and $1600 (250)$. The numbers in
the brackets are the size of the Floquet matrix ensemble. In
constructing the ensembles we have considered a range of
$\alpha$, i.e., $0.95 \leq \alpha \leq 1.05$. (b) Distributions
of $\ln (P_2)$ for the standard Dyson's COE matrices, with the same
matrix dimension as in the kicked top model and the same ensemble
size. (c) Analogous to Fig. \ref{Jx-JxDistAlphaJ1.0}(b) and Fig. 7(b), the scaling behavior of $\langle \ln (P_2) \rangle$
vs $\ln (J)$ is shown. Open circles are numerical results for the
kicked-top model, crosses are numerical results associated with
COE random matrices, and the solid curve represents the theoretical
prediction from the random matrix theory. The scaling shows that
$D_2=1$ in the standard kicked top model, which is dramatically
different from our observations made from the double-kicked top
model.}
\label{Haake_Top}
\end{figure}

Finally, as a numerical ``control" experiment, we study the
Floquet eigenstate statistics of the standard kicked top model.
This will help appreciate the difference between a normal driven
system and a critical driven system, both of which can have a
chaotic classical limit. Consider then the following Floquet
operator for the standard kicked top model \cite{KTM},
\begin{equation}
F_{\text{KTM}} = \exp\left(- i \frac{\eta J_z^2}{2 J}\right) \exp(-i \alpha J_x),
\end{equation}
which is just the last two factors of Eq. (\ref{Floquet_COE}),
with the same parity symmetry and time-reversal symmetry as the
$J_x-J_x$ model. In addition, we set the parameter
$\eta/J = \hbar_\eta$ at the same value as given in Eq.
(\ref{parameter1}). We construct a statistical ensemble by considering a range of $\alpha$, i.e. $0.95 \leq \alpha \leq 1.05$ (with chaotic classical limits).  We carry out the Floquet eigenstate statistics in the odd-parity subspace, whose dimension is $J$.  Because the classical limit is found to be chaotic, we compare the statistics with that associated with Dyson's COE matrices in random matrix theory (RMT).

Figure \ref{Haake_Top}(a) and (b) compare $\Pi[\ln (P_2)]$
associated with $F_{\text{KTM}}$ with that obtained from COE matrices, for different $J$. The difference between the actual dynamical system
and the COE can hardly be seen. Figure \ref{Haake_Top}(c) depicts
$\langle\ln (P_2)\rangle $ as a function of $\ln (J)$, with the
results of the kicked top (open circles) almost on top of those of
COE matrices (crosses). The solid line in Fig.
\ref{Haake_Top}(c) represents the theoretical curve for $\langle\ln
(P_2)\rangle $ obtained from RMT, i.e., $\langle \ln (P_2) \rangle
\sim \ln 3 - \ln (J)$.  The agreement between numerical COE results,
analytical RMT result, and the kicked top system as a classically
chaotic dynamical system is almost perfect. From the curve shown in
Fig. \ref{Haake_Top}(c), it is clear that $D_2$ here is unity and as
such the system does not show critical behavior. This non-critical
behavior indicates that the Floquet states of the kicked top model
are essentially random states, a feature fundamentally different
from our double-kicked top system that has a butterfly spectrum and
critical statistics in the Floquet eigenstates. It is also
interesting to note that in Fig. \ref{Haake_Top}(a) and (b), as $J$
increases, $\Pi[\ln (P_2)]$ becomes narrower and develops higher
peaks. This is an indication that, unlike the critical cases studied
above, $\Pi(\ln P_2)$ for the standard kicked top model approaches a
Dirac-delta type singular function with zero width (i.e.
$\sigma^2(\infty) \rightarrow 0$) as $J$ increases.

\section{\label{sec6} Concluding Remarks}

In this numerical study we have examined the statistics of the Floquet eigenstates of a recently proposed double-kicked top model with multifractal Floquet spectra. Following the methodologies used in studies of Anderson transition in time-independent systems, we have shown that the Floquet eigenstates associated with multifractal Floquet spectra also display critical behavior. In particular, we focus on the distribution of $\ln (P_2)$ and examine how the quantity $\langle \ln (P_2)\rangle$ averaged over all states
scales with the dimension of the Hilbert space $N$. It is shown
that $\langle \ln (P_2)\rangle $ scales linearly with $\ln (N)$,
with the slope of this linear scaling giving the fractal dimension
$D_2$ of the Floquet eigenstates.  The values of $D_2$ are found
to be far from unity (as a comparison, we showed that similar
analysis for a standard kicked top with a chaotic classical limit
yields $D_2=1$), constituting strong evidence that the Floquet
eigenstates are fractal and hence lying between localized and
delocalized states. Though we have worked on $P_2$ only, we note
that similar analysis can be done for $P_q$ defined in Eq.~(\ref{pq}).  One may then define a generalized fractal dimension
$D_q$ and further establish the multifractal nature of the
Floquet eigenstates.

The variance of $\ln (P_2)$, denoted $\sigma^2(N)$ for a Hilbert
space of dimension $N$, is also examined.  In Anderson-transition
studies with PRBM, $\sigma^2(N)$ is known to scale as
$N^{-\gamma}$ with $\gamma = D_2/(2\beta)$ for one-dimensional
systems, where $\beta = 1\,(2)$ for a system with (without)
time-reversal symmetry. By contrast, in our critical driven
system, $\sigma^2(N)$ is seen to scale similarly, but with $\gamma
= D_2/\beta$.  This reflects an interesting difference between
time-dependent systems and time-independent systems. Indeed,
eigenstates of PRBM are to model those of critical Hermitian
operators, whereas Floquet eigenstates of a critical driven system
should be understood in terms of critical unitary operators. To justify this
understanding, we have introduced a random unitary
matrix ensemble called PRBUM, with the variance of the matrix
elements of the unitary matrices following a power-law
distribution. We show that the eigenstates of PRBUM share many
critical statistical features with the double-kicked top model.
Most important, the variance of $\ln (P_2)$ of PRBUM does scale as
$N^{-(D_2/\beta)}$, which is the same as in the
double-kicked top model as a critical driven system.  We hence
anticipate that this scaling property of the variance of $\ln
(P_2)$ may be general in critical driven systems.  These results
complement the spectral results in Ref.
\cite{JiaoJiangbin2}
and should motivate further mathematical and theoretical studies in critical driven systems.


\section*{Acknowledgments}

J.W. acknowledges support from National Natural Science Foundation
of China (Grant No.10975115), and J.G. is supported by the NUS start-up fund (Grant
No. R-144-050-193-101/133) and the NUS ``YIA" (Grant No.
R-144-000-195-101), both from the National University of Singapore.

\appendix

\section{\label{App1} Mezzadri's algorithm}

This is a simple and numerically stable algorithm to generate the CUE matrices from
an ensemble of complex random matrices $\{Z_i\}$,  whose elements are Gaussian distributed
random numbers with mean zero and variance {\it unity}. In particular, applying the Gram-Schmmidt
ortho-normalization method to the columns of an arbitrary complex matrix $Z_i$, one can
factorize $Z_i$ as:
\begin{equation}
Z_i = Q_i R_i,
\end{equation}
where $Q_i$ is a unitary matrix and $R_i$ is an invertible upper-triangular matrix. One
can easily prove that the above factorization is not unique. Because of this
non-uniqueness, the random unitary matrices $\{Q_i\}$ are not distributed with Haar
measure \cite{Mezzadri}, i.e., the $\{Q_i\}$ matrices are not uniformly distributed over the
space of random unitary matrices. Fortunately, this factorization can still be made unique by imposing
a constraint on the $R_i$ matrices. By some group theoretical arguments, it was shown \cite{Mezzadri} that
if one finds a factorization such that the elements of main diagonal of $R_i$ become
real and strictly positive, then $\{Q_i\}$ matrices would be distributed with Haar
measure and hence form CUE.  Following these results,  the major steps of Mezzadri's algorithm are the following. First,
we start with an $N \times N$ complex Gaussian random matrix $Z_i$. Second, we factorize
$Z_i$ by any standard $QR-$decomposition routine such that
$Z_i = Q_i R_i$. Third, we create a diagonal matrix
\[\Lambda = \mbox{diag} \left(\frac{r_{11}}{|r_{11}|}, \dots,
\frac{r_{NN}}{|r_{NN}|}\right),\]
where $\{r_{ll}\}$ are the diagonal elements of $R_i$. As a final step, we
define $R_i^\prime \equiv \Lambda^{-1} R_i$ and $Q_i^\prime \equiv Q_i
\Lambda$. By construction, the diagonal elements of
$R_i^\prime$ are always real and strictly positive, and as such $\{Q_i^\prime\}$ would be
distributed with Haar measure and can be used to form the desired CUE.
The symmetric COE matrices can be constructed from the CUE matrices in a very simple
manner. In particular, let
$U$ be a member of the CUE generated above, then it can be shown that $V = U U^T$ will be a member of COE.
For the generation of PRBUM advocated in this work, we propose to replace $Z_i$ in the first step by a member in the PRBM ensemble that models
Anderson transition. Though there is no mathematical theory for our procedure, the uniformly distributed eigenphases (not shown here) of our PRBUM ensemble thus generated suggest its uniform distribution.

\end{document}